\documentclass[aps,pra,showpacs,twocolumn,superscriptaddress,superscriptaddress,]{revtex4-1}

\usepackage{graphicx}
\usepackage{color}% Include figure files
\usepackage{bm}% bold math
\usepackage{amssymb,amsmath}
\usepackage[dvipsnames]{xcolor}

\usepackage[caption=false]{subfig}

\begin{document}

\title{Excitation-assisted nonadiabatic charge-transfer reaction in a mixed atom-ion system}   

\author{Ming Li}
\affiliation{\em Department of Physics, Temple University, Philadelphia, PA 19122, USA}
\author{Michael Mills}
\author{Prateek Puri}
\affiliation{\em Department of Physics and Astronomy, University of California, Los Angeles, California 90095, USA}	
\author{Alexander Petrov}
\affiliation{\em Department of Physics, Temple University, Philadelphia, PA 19122, USA}
\affiliation{\em NRC Kurchatov Institute PNPI, Gatchina, Leningrad district 188300;
and Division of Quantum Mechanics, St. Petersburg State University,
St. Petersburg, 199034, RF}
\author{Eric R. Hudson}
\affiliation{Department of Physics and Astronomy, University of California, Los Angeles, California 90095, USA}
\author{Svetlana Kotochigova}
\affiliation{\em Department of Physics, Temple University, Philadelphia, PA 19122, USA}

\begin{abstract}
An important physical process unique to neutral-ion systems is the charge-transfer (CT) reaction. 
Here, we present measurements of and models for CT processes between co-trapped ultracold 
Ca atoms and Yb ions under well-controlled conditions. The theoretical analysis reveals the existence of three  
reaction mechanisms when lasers from a magneto-optical trap (MOT) and an additional catalyst laser
are present.
Besides the direct CT involving existent excited Ca population in the MOT,
the second pathway is controlled by MOT-induced CT, whereas 
the third one mostly involves the  additional red-detuned laser. 
\end{abstract}

\maketitle

\section{Introduction}
Over the last few decades the study of individual quantum
systems decoupled from external perturbations has become a reality.
Combining quantum-degenerate gases of fermionic or bosonic atoms,
held in electro-magnetic traps with a wide range of geometries,
with cooled and trapped ions is an exciting
and dynamic area in physics.  Cold and trapped atom-ion mixtures
can be engineered with a high level of control, detected state
selectively, and even constructed at the single-ion level.  
The majority of experimental and theoretical research into  
charge-transfer (CT) with ultracold atoms and ions has focused on their
collisions when prepared in their electronic ground state
\cite{Cote2000,Watanabe2002,Zhao2004,Grier2009,
Zhang2009,Idziaszek2009,Zipkes2010,Schmid2010,Hall2011,Liu2010,Zhang2011,Tacconi2011,Rellergert2011,
Lamb2012,Belyaev2012, Belyaev2013,McLaughlin2014,Haze2015}. 
Often, however, cold atom-ion experiments involve holding the
neutral atoms in a magneto-optical trap (MOT), providing opportunities
for scattering of electronically-excited atoms with the co-trapped
ions. Although the first steps towards understanding these collisions
have been reported \cite{Hall2011,Sillivan2012,Puri2017},
theoretical details are still poorly understood.

Charge-transfer can only be realized through transitions between
two or more potential energy surfaces (PESs)
that are characterized by electron transfer from  the neutral atom to
the ion, i.e. ${\rm A}^++{\rm B} \to {\rm A}+{\rm B}^+$. 
In the conventional Born-Oppenheimer (BO) adiabatic picture, such
transitions occur due to non-adiabatic coupling induced by the nuclear
motion in the initial and final electronic states \cite{Nikitin2006Drake}.
Usually, this coupling occurs in a small localized range of
inter-particle separations $R$, when electronic BO potentials of the
same symmetry have a so-called avoided crossing following the
Wigner-Witmer non-crossing rule. 

%Charge-transfer rate coefficients for heteronuclear atom-ion pairs 
%in their electronic ground states are very small, in most cases of 
%the order of $10^{-14}$ cm$^3$/s or below
%\cite{Zipkes2010,Zygelman14,Haze2015,Tomza2015a}, 
%as the corresponding potentials have a broad avoided crossing 
%and their coupling  is weak. On the other hand, our study provides 
%clear evidence in favor of the involvement of excited states 
%in the charge-transfer reaction and explains the observed four order 
%of magnitude larger rate coefficients for atoms in a MOT. 
%Here, the molecular structure is  complex and dense and
%thus provides  more opportunities of strong coupling between different
%reaction pathways.

%{\color{blue}
When an avoided crossing between entrance and exit BO potentials is broad
like the cases for many heteronuclear atom-ion pairs in their electronic
ground states, the charge-transfer rate coefficients are very small,
in most cases of the order on $10^{-14}$ cm$^3$/s or below\cite{Zipkes2010,Zygelman14,Haze2015,Tomza2015a}. On the other hand,
in a region where molecular structure is complex and potential
curves are dense, there is large probability of having narrower 
avoided crossings that will lead to much higher charge-transfer rate 
coefficients, sometimes approaching values of universal models
\cite{Gao11,Puri2017}. Our study provides clear evidence of such 
a situation when a number of closely lying excited potentials, populated 
through excitation laser, couple strongly via non-adiabatic couplings 
leading to significant charge-transfer reactions. Even though the atoms 
in the MOT spend most of their time in the ground electronic state, 
the rate coefficients can still reach four orders of magnitude higher
than that of pure ground-state CT reactions. 
%}

Here, we study collisions between Ca atoms in a MOT and 
Yb$^+$ ions in a co-located linear-quadrupole ion trap.
In this system, CT reactions
involve excited 4s4p$\,^1$P Ca atoms and ground-state Yb$^+$ ions.
Experimental and theoretical CT rate coefficients are
obtained and compared for temperatures $0.01\ {\rm K}< T < 1$ K.
Theoretically, we use hybrid quantum simulations based on ground- 
and excited-state molecular potentials that combine quantum 
close-coupling calculations with rate equations for populations.  
Special attention has been given to the long-range
induction and dispersion interactions  within the molecular complex
as the dissociation limits of excited-state potentials are in
proximity and non-adiabatic transfer between the potentials occurs at
relatively large separations. We incorporate  spontaneous emission from
excited-state potentials and include survival probabilities as an
important element in our model.   To further elucidate
the role of excited states in CT an additional catalyst
laser with a frequency that is red detuned from that of the MOT
laser is applied. As we will show, the effect of spontaneous emission
on the reaction path is then suppressed.

We will show that up to three  mechanisms  or pathways contribute 
to the reaction outcome. Only the third pathway involves 
the additional catalyst laser.  
In the first, an atom in the excited state collides and
reacts with the ion.  In the second, a ground-state atom
in the presence of the long-range interaction from a ground-state  
ion is resonantly excited by absorption of a photon from the MOT 
lasers and then reacts with the ion. Finally, for the third pathway 
a colliding ground-state atom-ion pair  absorbs a photon from the
tunable catalyst laser
%, is photo-associated to an excited-state potential,
and then reacts.  In all pathways the long-range
%, large separation
interaction potentials between the cold atom and ion together with 
spontaneous emission from the electronically-excited atom-ion complex
determine the rate coefficients.

%{\color{red} TO DO: The neutral Ca and Yb atom have similar electronic
%properties.  For example, their ionization energies and static
%polarizability of their ground state agree to within 2\%. How does
%this explain that the excited $^3$D state of Yb is degenerate with the $^1$P state
%of Ca?}

\section{Modeling charge-transfer pathways}
%\noindent
%{\bf Molecular complex and pathways.} 

\subsection{Molecular complex and pathways}

\begin{figure}[t]
\includegraphics[width=0.9\columnwidth,trim=0 55 0 0,clip]{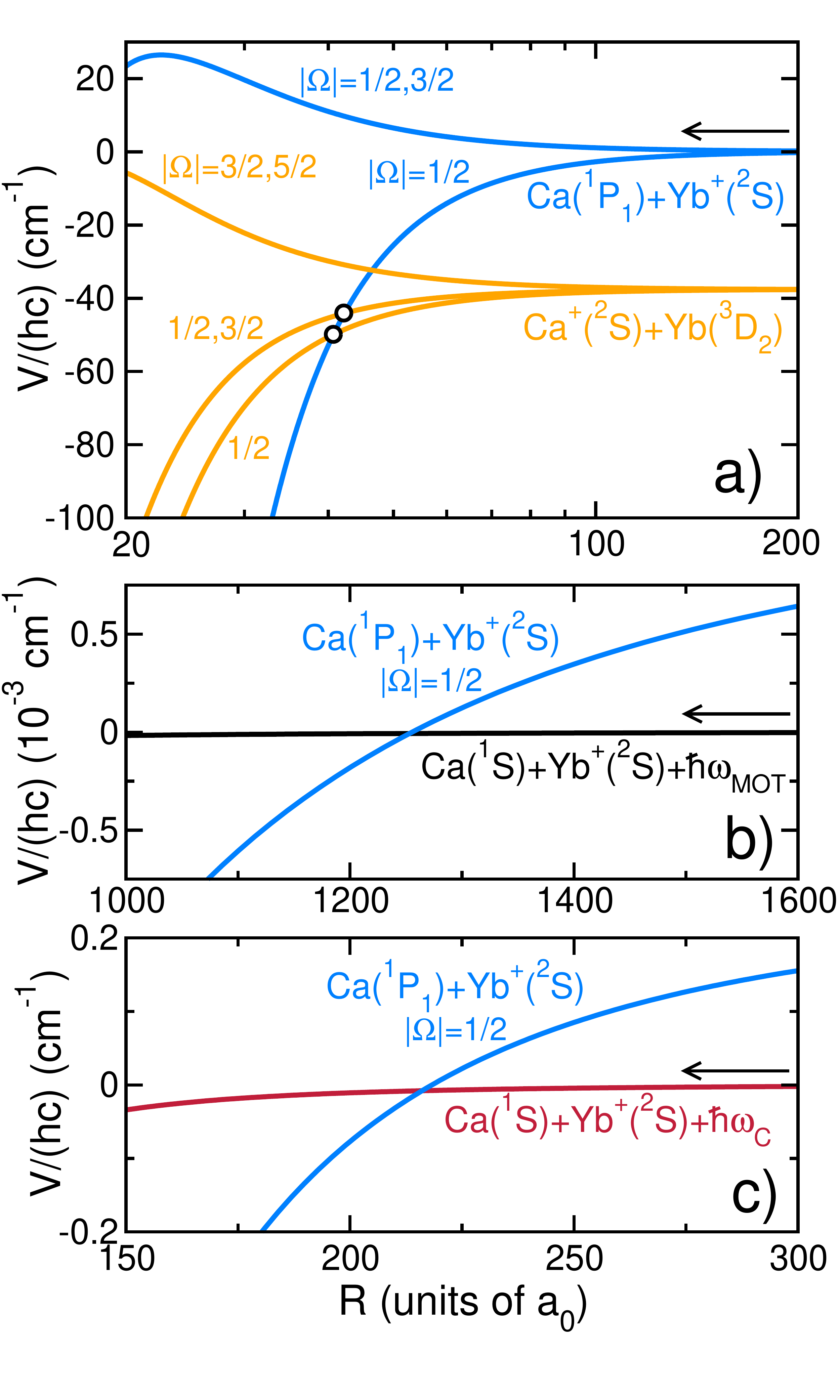}
\vspace*{-3mm}
\caption{a) Long-range  diabatic potential energy
curves in the atomic basis as   functions
of atom-ion separation $R$ on a logarithmic scale.   Blue and orange curves dissociate to
the Ca(4s4p\;$^1$P$_1$)+ Yb$^+$(6s\;$^2$S$_{1/2}$) and
Ca$^+$(4s\;$^2$S$_{1/2}$)+Yb(5d6s\;$^3$D$_2$)  limits, respectively.
Curves are labeled by  $|\Omega|$ and the zero of energy is located
at the top-most dissociation limit.  The two  crossings between
potentials relevant for  charge-transfer  are indicated with  
circular markers.  b) Photon-dressed  potential energies as
functions of $R$ for pathway II as defined in the text. The
black curve is the dressed-state potential dissociating to
Ca(4s$^2$\;$^1$S$_0$)+Yb$^+$(6s\;$^2$S$_{1/2}$) plus one MOT photon.
The blue curve is for the attractive potential dissociating to the
Ca(4s4p\;$^1$P$_1$)+Yb$^+$(6s\;$^2$S$_{1/2}$) limit.  c) Photon-dressed
potential energy curves as functions of $R$ for pathway III.
The dark red curve is the dressed-state potential dissociating to
Ca(4s$^2$\;$^1$S$_0$)+Yb$^+$(6s\;$^2$S$_{1/2}$) plus one catalyst
 photon.  The blue curve is as in panel b).  Black arrows
in panels a), b), and c) indicate the entrance channel for pathway
I, II, and III, respectively.  The zero of energy in panels b) and c)
is located at the dressed ground-state dissociation limit. }
\label{fig:PotLR}
\end{figure}

We start our analysis with the potential energy landscape for the
excited-state CT reaction.  Figure \ref{fig:PotLR}(a)
shows the relevant long-range diabatic $|\Omega|=1/2$, 3/2 and 5/2 potentials,
derived from the multipole expansion of the molecular forces,
dissociating to the Ca(4s4p$\,^1$P$_1$)+Yb$^+$(6s$\,^2$S) limit as
well as to the nearly-degenerate Ca$^+$(4s$\,^2$S)+Yb(5d6s$\,^3$D$_2$)
limit. Their asymptotic splitting is only $\Delta=hc\times 37.7$
cm$^{-1}$, where $h$ is the Planck constant and $c$ is the speed
of light in vacuum.  Moreover, the molecular electronic state of
each diabatic potential is a unique element of the separation-independent
``atomic basis'' of products of the relevant atomic or ionic Ca and
Yb states. The projection of the total electron angular momentum
on the internuclear axis, $\Omega$, is a good quantum number.  
Charge-transfer only occurs between states with the same $\Omega$, which
for our system occurs for $|\Omega|=1/2$ potentials near the two
crossings at $R\approx40a_0$, where $a_0$ is the Bohr
radius.  
Spin-orbit couplings are included which are essential for the exit
channels dissociating to the Ca$^+$(4s$\,^2$S)+Yb(5d6s$\,^3$D$_2$) limit.
Details of our calculation of the potentials and, in
particular, the evaluation of the strength of the coupling near the
crossing points can be found below as well as in Appendix \ref{app:pot}.

In a MOT, Ca is present in both its ground 4s$^2$ $^1$S$_0$ and
excited 4s4p $^1$P$_1$ state. We then define charge-transfer pathway I as
\[
{\rm Ca}(^1{\rm P}_1)+{\rm Yb}^+(^2{\rm S}_{1/2})  \to {\rm Ca}^+(^2{\rm S}_{1/2})+{\rm Yb}(^3{\rm D}_2) \,,
\]
where the initial state is indicated by the arrow in Fig.~\ref{fig:PotLR}(a)
and pathway II as
\[
{\rm Ca}(^1{\rm S}_0)+{\rm Yb}^+(^2{\rm S}_{1/2})+\hbar \omega_{\rm MOT} \to {\rm Ca}^+(^2{\rm S}_{1/2})+
{\rm Yb}(^3{\rm D}_2) \,.
\] 
This second pathway is assisted or dressed by a MOT photon with energy $\hbar \omega_{\rm
MOT}$ and the $|\Omega|=1/2$ Ca(4s4p $^1$P$_1$)+Yb$^+$(6s $^2$S$_{1/2}$)
potential is populated as an intermediate state, which then has
CT to Ca$^+$+Yb as in the first pathway.  Here, $\hbar$ is the 
reduced Planck constant. 
The MOT photon is detuned one natural linewidth, $\Gamma$,  to the red of the Ca $^1{\rm S}_0 \to ^1{\rm
P}_1$ transition leading to an avoided crossing at separations of more
than a thousand Bohr radii. The repulsive excited $|\Omega|=1/2$ and
$3/2$ channels are also populated due to the laser coupling, but do
not lead to significant CT reaction.

The third (III) pathway
\[
{\rm Ca}(^1{\rm S}_0)+{\rm Yb}^+(^2{\rm S}_{1/2})+\hbar \omega_{\rm C} \to {\rm Ca}^+(^2{\rm S}_{1/2})+
{\rm Yb}(^3{\rm D}_2) 
\]
is also photon assisted. In this case a tunable catalyst
laser with frequency $\omega_{\rm C}$  is introduced with the goal
to enhance the charge-transfer rate coefficient. The dressed
ground-state potential is shown in Fig.~\ref{fig:PotLR}(c) and
crosses the  same intermediate potential as in the second pathway.
Here, the diabatic crossing and coupling occur at
separation $R$ of  $200a_0$ to $500a_0$.  The laser
is detuned to the red of the Ca $^1{\rm S}_0 \to ^1{\rm P}_1$
transition by tens to hundreds of  $\Gamma$.

%
%\vspace*{0.2cm}
%\noindent
%{\bf Model ingredients.}

\subsection{Model ingredients}

Conventionally one would compute CT rate coefficients for
scattering from the potentials shown in Fig.~\ref{fig:PotLR}(a), their
$\Omega$-conserving electronic couplings, as well as $\Omega$-changing
(and conserving) couplings induced by the relative atom-ion rotational
interaction using a coupled-channels (CC) model.  For the relevant
collision energies $E=k_{\rm B}\times 1$ mK to $k_{\rm B}\times10$ K
and the long-range $1/R^3$ charge-quaduple nature of the
potentials, however, contributions to the rate coefficients from a large
number of total molecular angular momenta $\vec J$ need to be included.
Here, $k_{\rm B}$ is the Boltzmann constant and  $\vec J$ is the vector sum of the atom-ion total angular momenta
and the relative mechanical orbital angular momentum $\vec l$, which is
conserved in the absence of radiation.

To keep the computational effort tractable we employ the infinite-order
sudden approximation (IOSA)~\cite{Pack74,Secrest75,Hunter75,Kouri79},
in which Coriolis couplings between different $\Omega$ states are 
neglected and, for a given $J$, uses a centrifugal potential
$\hbar^2[\mathcal{L}(\mathcal{L}+1)]/(2\mu R^2)$ for each diagonal
matrix element of the potential matrix. Here the integer-valued
$\mathcal{L}$ is an ``average'' orbital angular momentum quantum
number and $\mu$ is the reduced mass. We choose $\mathcal{L}=J-1/2$
justified by the observation that for our entrance channels the sum
of the atom-ion total angular momenta is $1/2$ (in units of $\hbar$).
The resulting potential matrix is block diagonal in $\Omega$, $J$,
and the projection $M$ of $\vec J$ along the space-fixed laboratory
axis. In fact, the matrix and thus rate coefficients are independent
of $M$.  Consequently, we only need to solve for a small set of
coupled Schr\"odinger or coupled-channels (CC) equations for each
$J$ with $M=0$ and $\Omega=1/2$ using standard methods~\cite{Gordon69}.

Moreover, collisions on the four degenerate repulsive ${|\Omega|=1/2}$
and $3/2$ Ca(4s4p\;$^1$P$_1$)+Yb$^+$(6s\;$^2$S$_{1/2}$) potentials
will not lead to significant charge-transfer as the reactants for our
small relative collision energies are unlikely to tunnel through the
$\approx hc\times 20$ cm$^{-1}$ barrier of these repulsive potentials.
Nevertheless, these potentials will play an important role in the
rate coefficient as population in the corresponding states
is inevitable.

%
%\vspace*{0.5cm}
%\noindent
%{\bf Diabatic coupling.}

\subsection{Diabatic coupling}

Coupling between the diabatic channels is a second ingredient in
setting up our CC model. Its strength is most important where potentials
cross.  Figure \ref{fig:PotLR} shows three such points, but only two,
located at $R_c= 40.7 a_0$ and $ 42.3 a_0$, respectively, lead to 
charge-transfer. 
%{\color{blue} 
Their coupling, in the diabatic basis which stipulates that
the electronic wavefunctions barely change with $R$, comes from
the overlap between the wavefunctions of the transferring electron 
on either the Ca nucleus or the Yb nucleus. Such interaction is
Coulombic in nature and conserves the body-fixed
projection $\Omega$. Hence, only crossings between ${|\Omega|=1/2}$
potentials are relevant. The equivalent model in the adiabatic picture
would include an avoided crossing between BO potentials 
and a non-adiabatic coupling between them that mostly comes from
the d$/$d$R$ term in the Hamiltonian acting on the overlapping adiabatic
electronic wavefunctions.
%}

We construct a diabatic two-channel model~\cite{Nikitin2006Drake}
near each ${|\Omega|=1/2}$ crossing.  Since the two diabatic
basis functions have different electronic character, the 
corresponding non-adiabatic coupling in the adiabatic picture is 
localized and well approximated by a Lorentzian centered at the 
crossing point~\cite{Werner81}.
After transforming into the diabatic picture, we can write
$V_{12}(R)= [V_{11}(R)-V_{22}(R)] \tan [2\vartheta(R)]/2$,
where $V_{11}(R)$ and $V_{22}(R)$ are the two diabatic potentials
and  mixing angle $\vartheta(R)=\mathrm{arctan}[(R_c-R)/R_0]/2+\pi/4$
 with crossing location $R_c$ and coupling width $R_0$.
(With these definitions $V_{12}(R_c)\propto R_0$.)
The coupling width $R_0$ is taken to be the same for our two crossings
and will be adjusted to lead to theoretical rate coefficients
that agree with experimental values in cases where only the first two
pathways are involved. The resulting coupling width is used later for
all three pathways.

%\vspace*{0.3cm}
%\noindent
%{\bf Laser-induced coupling.}

\subsection{Laser-induced coupling}

The MOT and catalyst lasers couple the initial photon-dressed
Ca(4s$^2$\;$^1$S$_0$) + Yb$^+$(6s\;$^2$S$_{1/2})+\hbar\omega_{\rm MOT,C}$
and excited Ca(4s4p\;$^1$P$_1$) + Yb$^+$(6s\;$^2$S$_{1/2}$) channels.
Using a dressed-state approach~\cite{Petrov17} and in the IOSA we find
coupling matrix element $-(1/\sqrt{3})\,d\,\sqrt{I/(2c\epsilon_0)}$ in
SI units between the initial $\Omega=\pm 1/2$ channel and the attractive
$\Omega=\pm 1/2$ excited channel with the same $J$. 
%{\color{red}(The matrix element is the
%same between  $\Omega=-1/2$ to $-1/2$ channels.)} 
Here, $I$ is the MOT or catalyst laser intensity, 
$\epsilon_0$ is the electric constant, dipole moment
$d=\sqrt{S/3}=2.85ea_0$, using line strength $S$ of the 4s$^2\;^1{\rm
S}_0$ to 4s4p$\;^1{\rm P}_1$ transition of Ca~\cite{Yu18}, and $e$ is the
electron charge.  The factor $1/\sqrt{3}$ accounts for the polarization of
the laser projected onto the body-fixed coordinate frame. 
%In the IOSA without Coriolis couplings, a photon transition 
Direct laser-induced couplings
between the ground $\Omega=\pm 1/2$ and the attractive excited  
$\mp 1/2$ channels do not occur. 
This is because in the body-fixed frame, 
the attractive excited channels has $\Omega_\mathrm{Ca}=0$ and 
the transition preserves the projection quantum number of Yb due 
to the fact that the transition dipole moment
in the long-range (where the transition most likely to happen) 
originates from the excitation of outer electrons of the Ca atom.
The lasers also couple the ground-state channel to repulsive excited
$|\Omega|=1/2$ and $3/2$ 
channels. The repulsive channels, however, do not significantly contribute 
to the charge-transfer process and their laser-induced coupling matrix 
elements are not required.

Laser-induced couplings persist to $R\to\infty$ and for pathways II
and III we must diagonalize the asymptotic potential matrix and use its
eigenvectors and the average partial wave quantum number $\mathcal{L}$
to define a dressed scattering basis.  For pathway I, where light
does not dress states, the original atomic basis states can be used.
For each of the three pathways we can then solve the coupled-channels
equation for $|\Omega|=1/2$ and each $\mathcal{L}$ (or equivalently $J$)
and compute the charge-transfer cross section 
$\sigma_i(E,\mathcal{L})$
for $i$=I, II, and III and relative kinetic energy $E$ of the 
corresponding initial state. The multiplicity factor of 
$(2\mathcal{L}+1)$ is included in obtaining the cross section.

%
%\vspace*{0.3cm}
%\noindent
%{\bf Spontaneous decay}

\subsection{Spontaneous decay}

Charge-transfer involving the excited Ca(4s4p$^1$P$_1$) state is
affected by spontaneous emission
\cite{Gallagher89,Julienne92,Boesten93}, which limits the probability
of colliding particles to remain in the excited channel and reach
the diabatic crossing region near $R\approx 40a_0$, where CT
is most likely to occur. For our first pathway, Ca(4s4p $^1$P$_1$)
and Yb$^+$(6s$^2$S$_{1/2}$) start at very large $R$.
For the second and third pathways the excitation to the intermediate
Ca(4s4p$^1$P$_1$)+Yb$^+$(6s $^2$S$_{1/2}$) state is resonant at
separations where the energy of the $|\Omega|=1/2$ ground-state
potential plus the energy of a laser photon equals the attractive
$|\Omega|=1/2$ of the intermediate channel as shown in
Figs.~\ref{fig:PotLR}(b) and (c). This occurs at $R\approx 1200 a_0$
for pathway II and between $200a_0$ and $500a_0$ for pathway III depending on detuning. 
The classical time for the atom and ion
to be pulled together to separations near $R_c$ by the attractive
excited potentials at ultra-cold collision energies can
approach or exceed the $\tau=4.59$ ns Ca(4s4p $^1$P$_1$) lifetime.

We account for this spontaneous decay by computing the survival 
probability $p_i(E,\mathcal{L})$ to reach crossing points $R_c$ 
for initial collision energy $E$ and average partial wave $\mathcal{L}$
for each pathway $i$~\cite{Suominen94}. In essence, the probability 
is based on computing the collision time along classical trajectories
on the attractive excited Ca(4s4p $^1$P$_1$) + Yb$^+$(6s $^2$S$_{1/2}$)
$|\Omega|=1/2$ potential. More details are given in Appendix \ref{app:pot}.  
The cross section obtained from the CC calculation and the survival
probability are combined to define total CT rate coefficient
\begin{equation}
   k_i(E) = f_i \sum_{\mathcal{L}=0}^\infty 
   p_i(E,\mathcal{L})\,v_{\rm rel}\sigma_i(E,\mathcal{L}) 
   \label{eq:rate}
\end{equation}
for $i$=I, II, and III, where $v_{\rm rel}=\sqrt{2E/\mu}$ is the absolute value of the
relative velocity $\vec v_{\rm rel}$. The factor $f_i=\eta/3$, $1-\eta$, and $1-\eta$ 
for $i$=I, II, and III, respectively.
For pathway I  it accounts
for the fact that in a MOT a fraction $\eta$ of the Ca atoms is in
the exited state and that only the two (degenerate) attractive $|\Omega|=1/2$ Ca(4s4p
$^1$P$_1$) + Yb$^+$(6s $^2$S$_{1/2}$) channels out of the six excited states lead to charge-transfer. 
For pathway II and III $f_i$ is simply the fraction of Ca atoms in the ground state
as  both  initial states, $\Omega=\pm1/2$, equally contribute to the CT rate coefficient.
We use the MOT parameter in Ref.~\cite{JointPaper}, which leads to 
$\eta=0.092$, in our calculations.
As the ion temperature $T_{\rm i}$ is much larger than that of the
atoms, the relative velocity distribution in the center-of-mass
frame can be described by the three-dimensional
Gaussian probability distribution 
$P(\vec{v}_\mathrm{rel}) \propto \mathrm{exp}(-\mu
v_\mathrm{rel}^2/[2k_{\rm B} T_{\rm eff}])$ with effective temperature
$T_{\rm eff}= m_{\rm Ca}T_{\rm i}/M$, $M=m_{\rm Ca}+m_{\rm
Yb}$, and Ca and Yb$^+$ masses $m_{\rm Ca}$ and $m_{\rm Yb}$,
respectively. We use this distribution to thermally average the
charge-transfer rate coefficient.

\section{Results}

\subsection{Pathway I and II: MOT-induced charge-transfer}

\begin{figure}
\includegraphics[width=1\columnwidth,trim=0 0 10 60,clip]{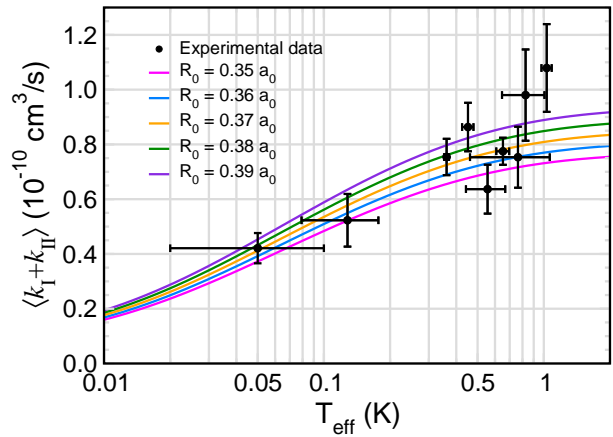}
\caption{Total thermalized charge-transfer rate coefficient in a Ca MOT
as a function of effective temperature $T_{\rm eff}$ in the
center-of-mass frame. Filled black circles with one-standard deviation
error bars are our experimental data points.
Solid lines are  theoretical predictions with coupling width $R_0$
ranging from $0.35 a_0$ to $0.39 a_0$.  The MOT laser has an intensity
of $78$ mW/cm$^2$ and is red detuned from the Ca $^1$S$_0$ to
$^1$P$_1$ transition by one natural linewidth, such that 9.2\% of the Ca atoms
are in the $^1$P$_1$ state.  }

\label{fig:mixparam}
\end{figure}

Figure~\ref{fig:mixparam} compares our total charge-transfer rate
coefficients as measured in the MOT  with the
thermalized theoretical  $\langle k_{\rm I}(E)+k_{\rm
II}(E)\rangle$ for several values of $R_0$ as a function of effective
temperature $T_{\rm eff}$ between 0.01 K and 2 K.  The data shows
a significant decrease of the rate coefficient due to the suppression
from spontaneous decays as the temperature lowers by over an order of
magnitude. Additional analysis shows
that about 40\% of the theoretical rate coefficient is due to the
first pathway. The figure also shows that at a fixed temperature
the rate coefficient increases monotonically when the coupling width
$R_0$ increases from $0.35 a_0$ to $0.39 a_0$. The theoretical
values agree well with the experimental data. The coupling strength
$V_{12}(R=R_c)$ for these $R_0$ at the crossing points is approximately
$hc\times 0.5$ cm$^{-1}$. In further support of the theoretical model 
and the resulting value of $R_0$, we obtained comparable coupling
strength at $R_c$ with a Heitler-London type of estimate~\cite{Tang98},
discussed in detail in Appendix \ref{app:pot}, using  
the overlap integral of atomic
orbitals and the electron-nuclei Coulomb potential.

\subsection{Pathway III: Photoassociation-enhanced charge-transfer}

\begin{figure}
 \includegraphics[width=0.9\columnwidth,trim=30 0 0 0,clip]{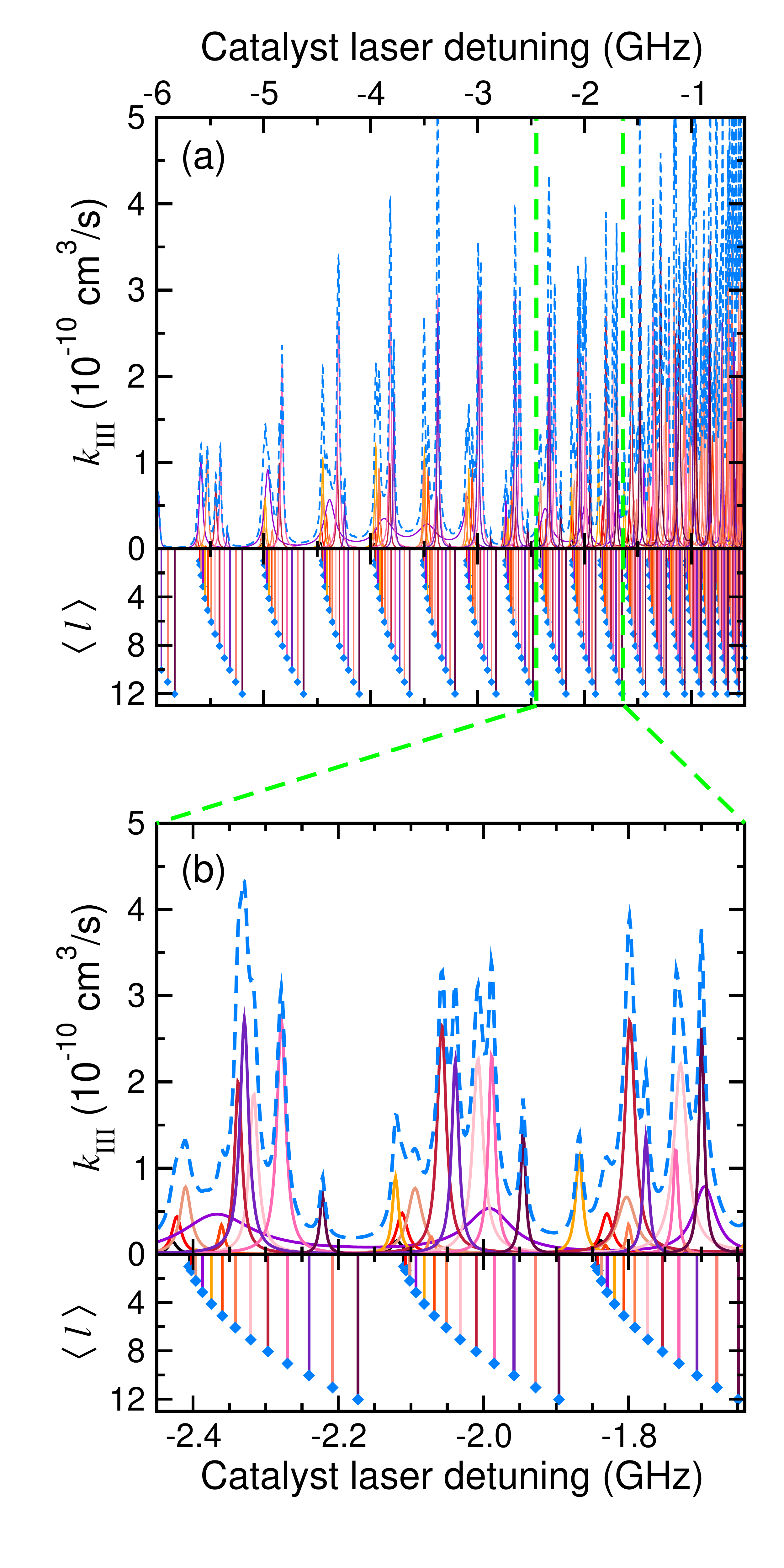}
 \caption{Charge-transfer rate coefficients  from the
catalyst laser and  assignment of catalyst 
resonances as functions of the detuning from the Ca
$^1$S$_0$ to $^1$P$_1$ transition for collision energy $E=k_{\rm B}\times1$ mK,
laser intensity of $I_{\rm C}=5$ W/cm$^2$, and
coupling width $R_0=0.37 a_0$.
The upper panel of subfigure a) shows rate  coefficients for 
detunings between $-6$ GHz and $-0.5$ GHz, while the upper panel of
subfigure b) shows a blowup near $-2$ GHz in order to better
distinguish the different curves. 
The dashed blue line corresponds to the total rate coefficient
from pathway III, while the various colored solid lines represent
contributions from average partial-wave channels  ${\cal L}=0$ to 12.
The lower panel in each subfigure shows the rovibrational $\Omega=1/2$ bound states dissociating to 
the Ca(4s4p\;$^1$P$_1$)+Yb$^+$(6s\;$^2$S$_{1/2}$) threshold.
The lowest thirteen rotational states for each vibrational
state are shown and the colors of the drop lines  mimic the
colors of the ${\cal L}$ contributions in the upper panels.
The $y$-axis of the lower panels is the expectation value of  $\vec{\cal L}$ of the bound states. }
\label{fig:kvsbnd}
\end{figure}

In the experiment described in Ref.~\cite{JointPaper} the addition
of a tunable laser with intensity $I_{\rm C}$ and (angular) frequency
$\omega_{\rm C}$ enhances the charge-transfer processes as the
third pathway is added. This laser is
detuned to the red of the Ca $^1{\rm S}_0$ to $^1{\rm P}_1$ transition
by tens to hundreds of natural linewidths. 
In fact, the laser excites rovibrational levels of the
attractive $|\Omega|=1/2$ Ca(4s$^2\;{^1{\rm P}_1})$+Yb$^+(6{\rm
s}\;^2{\rm S}_{1/2})$ potential. Since the MOT cooling lasers in
the experiment are always on, the third pathway coexists with the
other two. 

In this paper we want to highlight and focus on
the effect of the catalyst laser. The relevant diabatic
potentials are  shown in Figs.~\ref{fig:PotLR}(a) and (c).
The potential matrix is diagonalized asymptotically to
form the correct scattering basis.  By solving the CC equations,
we obtain the partial cross sections for the third pathway,
$\sigma_{\rm III}(E,\mathcal{L},\omega_{\rm pa})$.  
The survival probability $p_{\rm III}(E,\mathcal{L},\omega_{\rm C})$ 
is larger than for the second pathway. This is because the crossing 
between the dressed entrance channel and the intermediate excited 
channel occurs at $R\sim 200 - 500$ a$_0$, depending on 
$\omega_{\rm C}$, which is much smaller than for the second pathway. 
Thus, the reactants are quickly accelerated along the excited 
attractive $1/R^3$ potential and need much less time to reach $R_c$
to react.

The addition of pathway III via the catalyst laser enhances
the charge-transfer reaction by adding $k_{\rm III}(E)$
to the total rate coefficient $k_{\rm tot}(E)$, while leaving 
$k_{\rm I}(E)$ and $k_{\rm II}(E)$ unchanged to good approximation. Figure~\ref{fig:kvsbnd}
shows an example of $k_{\rm III}(E)$ as a function of the laser
detuning at collision energy $E=k_{\rm B}\times1$ mK. It is evident that the charge-transfer reaction
occurs in a resonant fashion with a larger number of narrow peaks.
%In addition, the height of the resonant features  decreases with increasing (negative) detuning.
The resonance locations
cluster and correspond to rotational progressions of the vibrational series
of the attractive $|\Omega|=1/2$ potential dissociating to the
Ca(4s4p\;$^1$P$_1$)+Yb$^+$(6s\;$^2$S$_{1/2}$) threshold. 
The height of the resonant features decrease with increasing 
(negative) detuning as the overlap of 
%the 1 mK continuum wavefunction of the entrance channel with the discrete 
%bound states decreases. 
resonances decrease due to increasing ro-vibrational spacing in the excited
potential.
%Bound-state wavefunctions corresponding
%to resonances with larger detuning have smaller radial extend. 

Scattering from thirteen partial waves ${\cal L}$ contribute
significantly  to the charge-transfer as the 1 mK collision energy roughly corresponds  to
the height of the centrifugal barrier for the ${\cal L}=12$ channel.
To  illustrate this, we compare the locations of the 
resonances with the ro-vibrational bound states of the attractive {\it diabatic} potential 
dissociating to the Ca(4s4p\;$^1$P$_1$)+Yb$^+$(6s\;$^2$S$_{1/2}$)
threshold in Fig.~\ref{fig:kvsbnd}. The figure also shows the expected value of ${\cal L}$ 
for each resonance.  In Fig.~\ref{fig:kvsbnd}(a), the locations of the 
onset of each group of resonances  closely follow the binding energies of ro-vibrational series. 

Figure~\ref{fig:kvsbnd}(b) shows a blowup of the spectrum for three vibrational 
levels  in Fig.~\ref{fig:kvsbnd}(a). We  see that the location of rotational states ${\cal L}$ 
does not directly follow the location of the corresponding resonances. The shifts are due 
to interferences with the charge-transferred  Ca$^+$($^2$S)+Yb($^3$D$_2$) exit channels
induced by the non-perturbative short-range couplings. 

\begin{figure}
\includegraphics[width=1\columnwidth,trim=0 0 0 -10,clip]{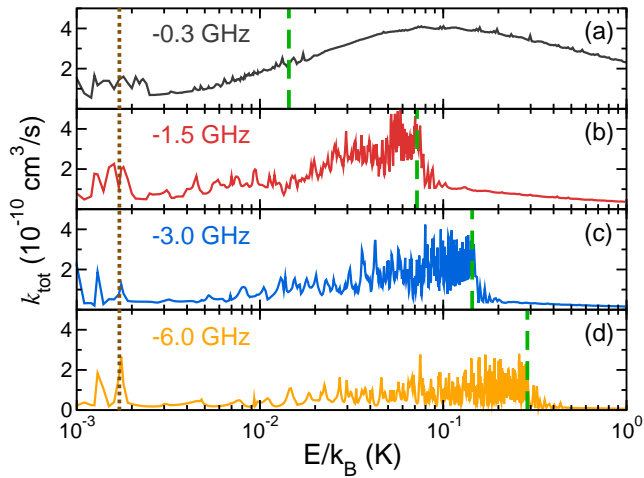}
\caption{Total charge-transfer rate coefficients in the presence of
the catalyst laser as functions of collision energy $E$.
In panels a), b), c), and d) the photo-association laser is detuned 
0.3~GHz, 1.5~GHz, 3.0~GHz, and 6.0~GHz 
to the red of Ca $^1$S$_0$ to $^1$P$_1$ transition, respectively. 
In each panel the  vertical 
dashed green line represents a collision energy equal to 
the energy-equivalent of the detuning of the catalyst laser.
The  vertical dotted brown line is located at  $E=h\Gamma=k_{\rm B}\times1.7$ mK,
both the energy equivalent of the MOT laser detuning and the natural linewidth of the Ca $^1$P$_1$ state.
We assume catalyst laser intensity $I_{\rm C}=6$ W/cm$^2$ 
and coupling width $R_0=0.39a_0$.  
  }
\label{fig:kvsT}
\end{figure}

Figure~\ref{fig:kvsT} shows an example total charge-transfer 
rate coefficient $k_{\rm tot}(E)$ as a function of collision energy 
at four detunings, $\delta$, of the catalyst laser.
The laser intensity is two orders of magnitude larger than $I_{\rm MOT}$. 
For  $\delta=-0.3$ GHz, $k_{\rm tot}(E)$  has a  
smooth behavior, interlaced with weak narrow features, and a maximum 
near $E=k_{\rm B}\times0.1$ K. For larger detunings sharp features 
dominate, while above a critical collision energy the rate coefficient 
rapidly approaches zero. For all detunings strong resonances are 
visible for ${E<k_{\rm B}\times 2}$ mK.

The behavior for large catalyst laser detunings 
can be understood from comparison of the collision energy 
with $h\delta$, the dashed green lines in Fig.~\ref{fig:kvsT}. 
For $E<h\delta$ pathway III contributes 
resonances to the total rate coefficient due to the coupling between 
the entrance channel continuum 
and the bound states of the attractive intermediate $|\Omega|=1/2$ 
potential.  When the collision energy 
exceeds  $h\delta$, the entrance continuum is only coupled to 
scattering states of the intermediate channel. Their  
coupling matrix elements are much smaller than
those between continuum-bound states and the rate coefficient 
becomes much smaller.

For  $\delta=-0.3$ GHz,  the total rate coefficient  
does not turn off at $h\delta$ thanks to pathway III 
as a consequence
of the fact that for small detunings the energy spacing between 
excitable vibrational levels  is smaller and 
resonances begin to overlap.  
In fact, helped by the relatively high-powered catalyst 
laser, interference between the broadened resonances
becomes important. Hence, the resonances behave almost like  
a continuum and the rate coefficient is 
a smooth function of $E$ both for energies smaller and larger 
than $h\delta$. The same effect is not obvious for the MOT 
laser pathway due to the much smaller laser intensity
that couples the continuum and
the bound states much weakly and does not broaden the
resonances nearly as much despite the smaller detuning.

Finally, pathway II contributes resonances to the total rate 
coefficient for collision energies comparable or smaller than 
$k_{\rm B}\times 2$ mK, roughly corresponding to  the detuning
of the MOT laser $h\Gamma=k_{\rm B}\times1.7$ mK labeled by
the brown dashed-lines. These resonances correspond to bound 
states with extremely-long-range outer turning 
points in the intermediate  $|\Omega|=1/2$ potential. 
%Because the outer turning points are so large and the intermediate 
%channel has a long-range attractive $1/R^3$ interaction, 
%absent for the entrance channel, the centrifugal barriers for the 
%intermediate state are not as high for the same $\cal L$. 
%As a results,  we find that many more partial waves contribute to 
%$k_{\rm II}(E)$ than to $k_{\rm III}(E)$. 

\section{Conclusion}
In conclusion, we have presented experimental measurements and results 
from a close-coupling model of photon-induced charge-transfer in 
Ca +Yb$^+$ that yielded insight into three contributing reaction 
mechanisms. The model relies on a dense manifold of 
electronically-excited long-range induction and dipolar potentials, 
their non-adiabatic coupling, and survival probabilities against 
spontaneous emission of the excited Ca atom. It leads to a high 
charge-transfer rate coefficient of the order of 10$^{-11}$ - 
10$^{-10}$ cm$^3$/s in agreement with the experimental results
\cite{JointPaper}.

\section{Acknowledgments}
Work at the Temple University is supported by the MURI Army Research Office Grant No. W911NF-14-1-0378 and 
Grant No. W911NF-17-1-0563, the U.S. Air Force Office of Scientific Research Grant No. FA9550-14-1-0321 
and the NSF Grant No. PHY-1619788.
Work at UCLA is supported in part by 
National Science Foundation 
(PHY-1255526, PHY-1415560, and DGE-1650604) and
Army Research Office 
(W911NF-15-1-0121, W911NF-14-1-0378, and W911NF-13-1-0213) grants.

\appendix 

\section{Long-range interaction potentials}
\label{app:pot}

We now describe in more detail the long-range diabatic interaction
potentials between excited Ca and Yb$^+$ coupled to Ca$^+$ and
excited Yb as shown in Fig.~\ref{fig:PotLR}(a).  The interaction
between an excited atom and an ion has two contributions.  The
first arises from the interaction between the ion charge and
the quadrupole moment of the excited atom and has an anisotropic
$C_3/R^3$ dependence on atom-ion separation $\vec R$, where $C_3$
depends on the orientation of $\vec R$. The second term is an
anisotropic $C_4/R^4$ interaction, where $C_4$ also depends on the
orientation of $\vec R$. It originates in second-order perturbation
theory from the interaction between the charge and the induced
dipole moment of the neutral atom.  Consequently, both $C_3$ and
$C_4$ only depend on the properties of the neutral atom.

Our diabatic potentials are the diagonal matrix elements of the
molecular interaction in the atomic basis in the body-fixed frame 
$|q_a j_a\Omega_a,q_bj_b \Omega_b\rangle=|q_a, j_a\Omega_a\rangle |q_b,
j_b\Omega_b\rangle$ labeled by charge state $q_s=0,+1$ and body-fixed
projection quantum number $\Omega_s$ of the angular momentum $j_s$
of the atom or ion along the internuclear axis, where $s=a$ and $b$
for Ca and Yb, respectively. This uniquely labels the atomic states 
relevant for our system. Electronic molecular
interactions always conserve $\Omega=\Omega_a+\Omega_b$.

Crucially, for our system both contributions to the long-range
potential are diagonal in this body-fixed basis \cite{Petrov17}.    
The matrix elements of $C_3$, expressed in two equivalent ways, are
\begin{eqnarray*}
C_{3,j_s\Omega_s}
 &=& q\frac{\langle j_s\Omega_s|j_s 2\Omega_s 0\rangle}
         {\langle j_s j_s|j_s 2 j_s 0\rangle} \frac{Q}{2} \\
 &=& q (-1)^{j_s-\Omega_s}
\begin{pmatrix}
       j_s & 2 & j_s \\
      -\Omega_s & 0 & \Omega_s
       \end{pmatrix}
       \langle j_s ||Q_2|| j_s \rangle\,,
\end{eqnarray*}
where the quantum numbers $j_s\Omega_s$ always describe the state of the neutral atom,
$q=+1$ for the corresponding ion, (:::) denotes a Wigner 3-$j$
symbol, and $\langle j_1 m_1|j_2 j_3 m_2 m_3\rangle$ is a Clebsch-Gordan
coefficient. Finally, $Q$ is the quadrupole moment defined in
Refs.~\cite{Mitroy08,Derevianko01}, while $\langle j_s ||Q_2|| j_s
\rangle$ is the reduced matrix element used by
Refs.~\cite{Buchachenko11,Derevianko01}.  For $^1$P$_1$ Ca state the quadrupole moment is positive with 
$|Q|=11.04 e a_0^2$ \cite{Mitroy08}. The sign convention is derived from Ref.~\cite{Ceraulo91}.
For  the $^3$D$_2$ state of Yb, $\langle j_s ||Q_2|| j_s
\rangle=+14.2 e a_0^2$~\cite{Buchachenko11}.

The diagonal matrix elements of the $C_4$ coefficient are~\cite{Mitroy10}
\[
C_{4,j_s\Omega_s}=-\frac{q^2}{2}\left[\alpha_{0,j_s}
  +\alpha_{2,j_s}\frac{3\Omega_s^2-j_s(j_s+1)}{j_s(2j_s-1)}\right] \,,
\]
where $\alpha_{0,j_s}$ is the static scalar polarizability and $\alpha_{2,j_s}$
is the static tensor polarizability of the neutral atom in 
state $|0,j_s\Omega_s\rangle$. For the Ca $^1$P$_1$ state, $\alpha_{0,1}=242.4 a_0^3$ and 
$\alpha_{2,1}=-55.3 a_0^3$~\cite{Mitroy08}.  For the Yb $^3$D$_2$ state,  $\alpha_{0,2}=61 a_0^3$ and 
$\alpha_{2,2}=28 a_0^3$~\cite{Buchachenko11,Bowers99}.

Finally, the long-range interaction between a neutral S-state atom and a
S-state ion has an isotropic, attractive $C_4/R^4$ dependence on $R$ . For Ca+Yb$^+$ it 
is shown in Figs.\ref{fig:PotLR}(b) and (c) as the
dressed state potential.  The $C_4$ coefficient equals $-\alpha_{0,0}/2$,
where  $\alpha_{0,0}$ is the static polarizability of the neutral atom.
The $C_3$ coefficient is zero as  S-state atoms have zero quadrupole moment.

Table \ref{tab:LR} gives the relevant $C_3$ and $C_4$ coefficients
as well as lists the quantum numbers of the channels. A negative
sign indicates attractive interactions.  At smaller separations
(not shown in Fig.~\ref{fig:PotLR}) each potential transitions to
a repulsive $C_{12}/R^{12}$  potential. 
%The dependence of rate
%coefficients on the short-range  potentials is discussed in main body of
%the article
%{\color{red} CHECK. Is this true? Should we give values for $C_{12}$?}.

\begin{table}
\caption{The  $C_3$ and $C_4$ coefficients of the attractive
long-range ${\Omega=1/2}$ diabatic potentials in atomic units and
quantum numbers for the corresponding Ca+Yb$^+$ or Ca$^+$+Yb channels.
Channels are uniquely described by the charge $q_i$, atomic angular
momenta $j_i$, and its body-fixed projection $\Omega_i$ on the
internuclear axis with $i=a$ and $b$ for Ca and Yb, respectively.
Potentials are degenerate for $-\Omega$ and $\Omega$, where
$\Omega=\Omega_a+\Omega_b$.}
\label{tab:LR}

\begin{tabular}{ccc|ccc|cc}
		\multicolumn{3}{c|}{Ca/Ca$^+$}  &  \multicolumn{3}{c|}{Yb/Yb$^+$}  \\
	  $q_a$ & $j_a$ & $\Omega_a$ & 
		$q_b$ & $j_b$ & $\Omega_b$ & $C_3$ & $C_4$\\
	\hline
	0 & 0 & 0   &  +1 & 1/2 & $1/2$ &  0 & $-78.55$\\
	0 & 1 & 0   & +1  & 1/2 & 1/2 &  $-11.04$ & $-176.74$\\
	+1 & 1/2 & 1/2  & 0 & 2 & 0 & $-3.39$ & $-16.5$\\
	+1 & 1/2 & $-$1/2  & 0 & 2 & 1 & $-1.70$ & $-23.5$\\
	\end{tabular}
\end{table}

Diabatic potentials with the same $\Omega$  cross and couple near 
${R_c\approx 40a_0}$. As discussed in the main text, we have opted to 
use model coupling function with coupling width $R_0$. The value
of $R_0$ is fitted to experimental data and estimated to be
between $0.35$ and $0.39$ a$_0$. In support of our model and fitting
result, we can also estimate the diabatic coupling strength at ${R=R_c}$
based on a Heitler-London method. In atomic units, we can write the 
coupling matrix element between the attractive $|\Omega|=1/2$
Ca(4s4p\;$^1$P$_1$)+ Yb$^+$(6s\;$^2$S$_{1/2}$) and  
Ca$^+$(4s\;$^2$S$_{1/2}$)+Yb(5d6s\;$^3$D$_2$) channels as  
\begin{equation}
V_{12} (R)\sim \langle {\rm Ca}(4p)| \frac{1}{r_{\rm Ca}}
  +\frac{1}{r_{\rm Yb}} | {\rm Yb}(5d) \rangle \,,
  \label{eq:Vex}
\end{equation}
where $|{\rm Ca}(4p)\rangle$ and $|\mathrm{Yb}(5d)\rangle$ are the Ca  $4p$  and Yb  $5d$ 
Hartree-Fock electronic   orbitals, respectively, and the electron coordinate  for the two orbitals is  
 $\vec{r}_{\rm Ca, Yb}$ with respect to the Ca and  Yb nuclei, respectively.
At $R=R_c$ we find that $V_{12} (R_c)$ is on the order of $hc\times 0.5$ cm$^{-1}$ which corresponds to the range of $R_0$ we obtained.

\section{Survival probabilities}
\label{app:survival}

The evaluation of survival probabilities within the IOSA framework on the attractive
excited potential $V_e(R)$ of the $|\Omega|=1/2$ Ca(4s$^2\;{^1{\rm
P}_1})$+Yb$^+(6{\rm s}\;^2{\rm S}_{1/2})$  channel  due to spontaneous
decay of the Ca ${^1{\rm P}_1}$ state can be treated with rate
equations for populations derived from the optical Bloch equations
\cite{OBECohenTannoudji,Suominen94}.  Here, the atom pair decays back to
the ground-state potential $V_g(R)$ of the $|\Omega|=1/2$ Ca(4s$^2\;{^1{\rm
S}_0})$+Yb$^+(6{\rm s}\;^2{\rm S}_{1/2})$  channel.
In a MOT we can assume that the coherence 
between the Ca ${^1{\rm S}_0}$ and ${^1{\rm P}_1}$ states decays much faster 
than those of the populations. Moreover, at our temperatures where a large number of
relative orbital angular momenta $\cal L$ contribute, the relative nuclear 
motion for the purpose of estimating the survival probability can be
described by classical evolution $R(t)$  on the potential
$U_e(R;\mathcal{L})=V_e(R)+\hbar^2\mathcal{L}(\mathcal{L}+1)/(2\mu R^2)$ 
from the excitation region at (very) large separation at $t=0$ to
$R_c\approx 40a_0$, the separation where charge-transfer occurs. 
At $t=0$ the atom pair has relative kinetic energy 
$E$ and is moving towards smaller $R$.

Under these assumption we have for  pathway I
\begin{equation}
    \frac{\mathrm{d}p_{e}(t)}{\mathrm{d}t}=-\Gamma p_{e}(t)
	-\Gamma'(t)\,p_e(t)+\frac{1}{3}\Gamma'(t)\, p_g(t)
 \label{eq:stimg}
\end{equation}
and ${p_g(t)+3p_e(t)=1}$, where $p_g(t)$ and $p_e(t)$ are the 
populations in the  ground- and excited-state channel, respectively.
%{\color{blue} We need something about the factor 3 in the populations.}
Here, $\Gamma$ is the natural linewidth of Ca(${^1{\rm P}_1}$), and
\begin{equation}
   \Gamma'(t) = A_\Gamma \frac{\gamma^3}{\Delta E(t)^2+\gamma^2} \;,
\end{equation}
describes the stimulated absorption and  emission rate of  MOT
photons, where  
%$\gamma$ is the fast dephasing rate of the coherence and
$\gamma=\Gamma/2$ and
the time-dependent $\Delta E(t)=V_e(R(t))-V_g(R(t))-\hbar\omega_{\rm MOT}$ 
at separation $R(t)$. The last term on the right-hand side of  
Eq.~\ref{eq:stimg} accounts for processes where, after a spontaneous 
emission event, the atom pair is  again excited and participates in 
the charge-transfer collision. The factor of one third in this term 
accounts for the fact that only one third of the photons 
are able to excite the system back to the attractive excited channels. 

The constant $A_\Gamma$ is set such that the steady-state
solution of Eq.~\ref{eq:stimg} for $R\to\infty$ reproduces the 
experimental fraction of atom pairs in the $|\Omega|=1/2$ excited 
potential $V_e(R)$, i.e. $p_e|_{R\to\infty}=\eta/3$, where $\eta$
is the fraction of Ca atoms in the $^1$P$_1$ state. For the
MOT parameters in Ref.~\cite{JointPaper} 
$\Delta E\to-\Gamma$
for $ R\to\infty$ and $\eta=0.092$.
%For simplicity, we have assumed $\gamma=\Gamma/2$.

In practice, we do not solve Eq.~\ref{eq:stimg} directly but rephrase
the  equation into one for separation $R$
by noting that $\mathrm{d}t= \mathrm{d}R/v(R;E,\mathcal{L})$, where 
velocity $v(R;E,\mathcal{L})$  satisfies 
$\mu v^2/2+U_e(R;\mathcal{L})=E$ for each $R$. The radial differential
equation can be integrated from very large $R$ to crossing point $R_c$ to
obtain survival probability $p_{\rm I}(E,\mathcal{L})$ for pathway I.

Our second pathway is also affected by spontaneous decay of 
the excited channels. In this case the excitation  occurs 
near $R_x\approx1200 a_0$. The stimulated excitation and decay 
are already included in the close-coupling calculations when the light
coupling is included and the asymptotic basis functions diagonalized, 
thus do not need to be included here.
We then find the simpler differential equation
\begin{equation}
	\frac{\mathrm{d}p_{e}(t)}{\mathrm{d}t}=-\Gamma p_e(t) \,,
\end{equation}
which is transformed into one for  $R$ and solved from $R=R_x$ with $p_e(R_x)=1$ 
to $R_c$ assuming an initial kinetic energy $E$ and average partial 
wave $\mathcal{L}$. The final value at $R_c$ defines the survival probability 
$p_{\rm II}(E,\mathcal{L})$ for this pathway. 
The differential equation for the third pathway is the same as for
pathway II, but now the excitation separation is even smaller
and we find that the $p_{\rm III}(E,\mathcal{L})$ are larger than $0.1$
for the detunings considered here.

\bibliography{CaYb+ML}

\end{document}